# An ansatz for the nonlinear Demkov-Kunike problem for cold molecule formation

H. H. Azizbekyan

*Institute for Physical Research of NAS of Armenia, Ashtarak, Armenia*
*Moscow Institute of Physics and Technology, Dolgoprudny, Russia*
*LPMC, Université Paul Verlaine – Metz, Metz, France*

**Abstract.** We study nonlinear mean-field dynamics of ultracold molecule formation in the case when the external field configuration is defined by the level-crossing Demkov-Kunike model, characterized by a bell-shaped coupling and finite variation of the detuning. Analyzing the fast sweep rate regime of the strong interaction limit, which models a situation when the peak value of the coupling is large enough and the resonance crossing is sufficiently fast, we construct a highly accurate ansatz to describe the temporal dynamics of the molecule formation in the mentioned interaction regime. The absolute error of the constructed approximation is less than $3 \times 10^{-6}$ for the final transition probability while at certain time points it might increase up to $10^{-3}$. Examining the role of the different terms in the constructed approximation, we prove that in the fast sweep rate regime of the strong interaction limit the temporal dynamics of the atom-molecule conversion effectively consists of the process of resonance crossing, which is governed by a nonlinear equation, followed by atom-molecular coherent oscillations which are basically described by a solution of the linear problem, associated with the considered nonlinear one.

**PACS numbers:** 03.75.Nt Other Bose-Einstein condensation phenomena, 33.80.Be Level crossing and optical pumping, 34.50.Rk Laser-modified scattering and reactions

## 1. Introduction

After the realization of Bose-Einstein condensation (BEC) [1] in dilute gases of neutral alkali-metal atoms, the next challenge was to achieve a BEC of molecules. Molecules have complex internal structure and, therefore, more degrees of freedom, thus they offer a vast range of properties not available in the case of atoms. However, in contrast to atomic condensates, achieving molecular condensates via standard laser cooling techniques [2] is practically impossible, because of the rovibrational spectrum of the molecules. Consequently, the Raman photoassociation [3] and magnetic Feshbach resonance [4] have become the standard tools to create cold molecules starting from ultracold atomic gases [5-7].

So far, the theoretical efforts to describe the dynamics of cold molecule formation (e.g., see Refs. [8-20]) have mostly been focused on the treatment of the Landau-Zener (LZ) model [21]. In the case of photoassociation the LZ model describes a situation when the two quantum states are coupled by an external optical field of constant amplitude and a variable frequency, which is linearly changed in time. But this situation has some drawbacks; it is unrealistic to have a constant coupling that never turns off or infinite energies at $t \to \pm\infty$. However, there exists a model that has all the virtues of the LZ model and is free from its shortcomings. Such a model is the first Demkov-Kunike (DK) quasi-linear level-crossing model of a bell-shaped pulse (vanishing at $t \to \pm\infty$) and finite detuning [22]. The DK model can be considered as a *physical generalization* of the LZ model.

In the present study we investigate the temporal dynamics of coherent molecule formation via photo- or Feshbach-association of ultracold atoms (under the conditions considered in the present research the two techniques, the photoassociation and Feshbach resonance, are mathematically equivalent; for definiteness, in what follows we use the photoassociation terminology). We consider a condensate initially being in all-atomic state since under contemporary experimental conditions one faces this case most frequently.

The weak interaction limit of the nonlinear DK problem, corresponding to small values of the peak laser field amplitude, has recently been discussed in Ref. [23], where an analytical formula describing the temporal dynamics of the molecule formation has been obtained. The strong interaction limit of the DK model, corresponding to large values of the peak laser field amplitude, has been studied in Refs. [24-26]. In Refs. [24, 25] it has been shown that the strong interaction limit of the DK problem is effectively subdivided into two different interaction regimes corresponding to slow and fast sweeps of the detuning through the resonance. When the passage through the resonance is slow, the system exhibits large-amplitude Rabi-type oscillations between atomic and molecular populations. In the opposite limit, in the case of the fast enough resonance crossing, only weak, damped oscillations between the atomic and molecular subsystems occur.

In Ref. [26], an approximate solution to the nonlinear DK problem in the large sweep rate regime of the strong interaction limit has been constructed. This approximation, defined as a solution of a first-order nonlinear equation, contains a fitting parameter which has been determined through a variational procedure. However, the constructed approximation misses several essential features of the association process such as the coherent oscillations between atomic and molecular populations which arise after the system has passed through the resonance. In the present development, we use the solution presented in Ref. [26] as a zero-order approximation to construct the next approximation term to the problem. The resultant approximation contains fitting parameters that we determine numerically. The numerical simulations show that the absolute error of the constructed formula is less than $3 \cdot 10^{-6}$ for the final transition probability while at certain time points it might increase up to $10^{-3}$.

## 2. Mathematical treatment

In the mean field two-mode approximation, the coherent dynamics of ultracold diatomic molecule formation is described by a basic semiclassical time-dependent nonlinear two-state model [27, 28]:

$$i\frac{da_1}{dt} = U(t)e^{-i\delta(t)}\bar{a}_1 a_2,$$
$$i\frac{da_2}{dt} = \frac{U(t)}{2} e^{i\delta(t)} a_1 a_1,$$
(1)

where $t$ is time, $a_1$ and $a_2$ are referred to as the atomic and molecular state probability amplitudes, respectively, $\bar{a}_1$ denotes the complex conjugate of $a_1$, the real function $U(t)$, referred to as the Rabi frequency, is proportional to the laser field amplitude of the associating field, and the real function $\delta(t)$ is the integral of the associated frequency detuning $\delta_t$ (hereafter, the minuscule alphabetical index denotes differentiation with respect to the corresponding variable). System (1) possesses the first integral $|a_1|^2 + 2|a_2|^2 = \text{const} = 1$ that manifests the particle number conservation during the interaction process. Since we consider the basic situation when the system starts from the all-atomic state, the initial conditions imposed are: $|a_1(-\infty)| = 1$, $a_2(-\infty) = 0$.

The external field configuration we discuss here is the DK model defined as

$$U = U_0 \text{sech}(t/\tau), \quad \delta_t = 2\delta_0 \tanh(t/\tau), \quad (\tau > 0).$$
(2)

where $\tau$ is a (positive) scaling parameter. In what follows we take $\tau = 1$ that is equivalent to the rescaling of the time (see Fig1).

Following the general scheme presented in Refs. [23-26], we first apply to the basic set of equations (1) the transformation of the independent variable

$$z(t) = \int_0^t \frac{U(t')}{U_0} dt',$$
(3)

reducing the equations to a constant-amplitude form. Further, we write an exact equation obeyed by the function $p = |a_2|^2$:

$$p_{zzz} - \frac{\delta_{zz}^*}{\delta_z^*} p_{zz} + \left[\delta_z^{*2} + 4\lambda(1-3p)\right]p_z + \frac{\lambda}{2}\frac{\delta_{zz}^*}{\delta_z^*}\left(1 - 8p + 12p^2\right) = 0,$$
(4)

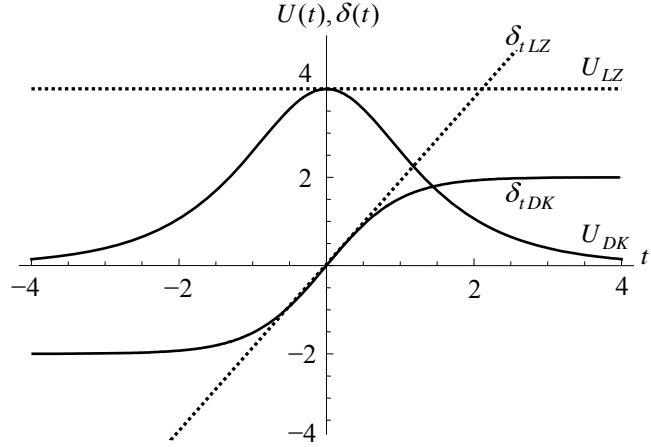

Fig. 1. Time dependence of $U(t)$ and $\delta(t)$. Solid curves – the Demkov–Kunike model: $U = U_0 \text{sech}(t)$, $\delta_t = 2\delta_0 \tanh(t)$, dotted lines – the Landau–Zener model: $U = U_0$, $\delta_t = 2\delta_0 t$.

where the function $\delta_z^*$, referred to as the effective detuning, is defined as

$$\delta_z^*(z(t)) = \delta_t(t) \frac{\sqrt{\lambda}}{U(t)} \tag{5}$$

with $\lambda = U_0^2$. Taking into account the definition of the DK model (2), we obtain $\delta_z^*(z(t)) = 2\delta_0 \sinh(t)$. Though the maximal value attained by the function $p$ is equal to $1/2$, $p \in [0, 1/2]$, we conventionally refer to $p$ as to the molecular state probability. For gaining a better intuitive understanding of the molecule formation in the large sweep rate regime of the strong interaction limit ($\lambda > 1$, $1 < \delta_0 < \sqrt{\lambda}$), in Fig. 2 we show the numerical plot of the molecular state probability in the mentioned interaction regime as a function of time.

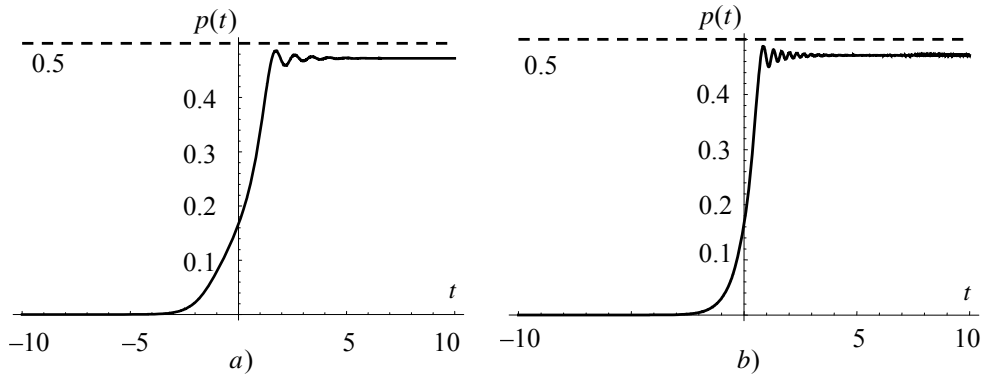

Fig. 2. The probability of the molecular state versus time for the DK model in the large sweep rate regime of the strong interaction limit. a) $\lambda = 100$, $\delta_0 = 4$; b) $\lambda = 100$, $\delta_0 = 10$.

In Ref. [26] the approximate solution of the exact equation for the molecular state probability (4) has been constructed by neglecting the two higher order derivate terms and adding to the truncated equation a term of the form $A\delta_{zz}^*/\delta_z^*$, where $A$ is an adjustable parameter. Thus, the zero-order approximation to the problem has been chosen as a solution of the following nonlinear equation of the first order:

$$\left[\delta_z^{*2} + 4\lambda(1 - 3p_0)\right] p_{0z} + \frac{\lambda}{2} \frac{\delta_{zz}^*}{\delta_z^*} \left(1 - 8p_0 + 12p_0^2\right) - A \frac{\delta_{zz}^*}{\delta_z^*} = 0. \tag{6}$$

The arguments to construct this approximation have been based on the fact that in the large sweep rate regime of the strong interaction limit the parameters $\lambda$ and $\delta_0$ are supposed to be large. As it has been shown in Ref. [26], the exact solution of the augmented limit equation (6), satisfying the imposed initial condition $p_0(t = -\infty) = 0$, is given as a solution of the following polynomial equation of the fourth order:

$$\frac{\lambda}{\delta_z^{*2}(z(t))} = \frac{p_0(p_0 - \beta_1)(p_0 - \beta_2)}{9(p_0 - \alpha_1)^2(p_0 - \alpha_2)^2}, \quad (7)$$

where

$$\alpha_{1,2} = \frac{1}{3} \mp \frac{1}{6}\sqrt{1 + \frac{6A}{\lambda}}, \quad \beta_{1,2} = \frac{1}{2} \mp \sqrt{\frac{A}{2\lambda}}. \quad (8)$$

Eq. [7] defines a *quartic algebraic equation* for determination of $p_0(t)$. Note that

$$p_0(0) = \alpha_1 \quad \text{and} \quad p_0(+\infty) = \beta_1. \quad (9)$$

The limit solution $p_0(z(t), A)$ is a monotonically increasing function that starts from zero at $t = -\infty$, reaches some value less than $1/6$ at $t = 0$ and tends to a finite positive value less than $1/2$ for $t \to +\infty$ when $0 < A < \lambda/2$ (see Fig. 3). In Ref. [26], an analytical expression for the parameter $A$ has been suggested. However, in the present development we do not specify the value of $A$ to consider it as a fitting parameter.

To proceed further, we now try to construct the first-order approximation to the problem using the limit function $p_0$ as a zero-order approximation. To do that, we make the change of the dependent variable

$$p = p_0 + u \quad (10)$$

in the exact equation for the molecular state probability (4). This transformation leads to the following exact equation for the correction term $u$:

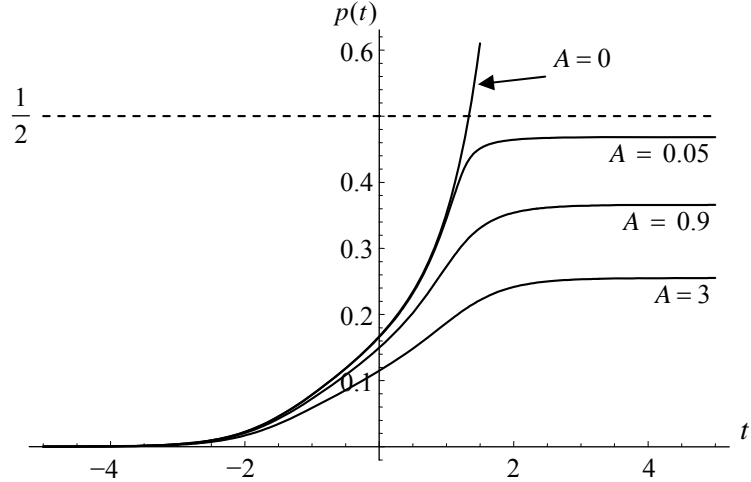

Fig. 3. The limit solution $p_0(t)$ (Eqs. (6)-(7)) vs. time for $A = 0, 0.05, 0.9, 3$ and $\lambda = 25$.

$$\left(\frac{d}{dz} - \frac{\delta_{zz}^*}{\delta_z^*}\right)\left(u_{zz} + 4\lambda(1 - 3p_0)u + p_{0zz} - A - 6\lambda u^2\right) - \delta_z^{*2} u_z = 0. \quad (11)$$

Since the function $p_0$ is supposed to be a good approximation for the molecular state probability $p$, the correction $u$ is supposed to be small. Thus, if we neglect the nonlinear term $-6\lambda u^2$, the exact equation (11) for $u$ will be linearized. By comparing the resultant linear equation with that obeyed by the second state probability $P_{DK}$, calculated within the linear theory of nonadiabatic transitions (for a detailed explanation see [23]), we see that if we consider $p_0$ as a constant, the solution of the constructed linearized equation will be given as a scaled solution to the *linear* DK problem with modified parameters. This observation gives an argument to make a conjecture that the exact solution of Eq. (11) can be approximated as

$$u = C^* \frac{P_{DK}(\lambda^*, \delta_0^*, t - t_{ph})}{P_{DK}(\lambda^*, \delta_0^*, \infty)}, \tag{12}$$

where $P_{DK}(\lambda^*, \delta_0^*, t)$ is the solution to the linear DK problem with effective parameters $\lambda^*$, $\delta_0^*$, and $t_{ph}$ being an extra temporal shift.

By combining Eqs. (10) and (12) we arrive at the following principal conjecture: an accurate approximation describing the time evolution of the molecular state probability can be written as a sum of the solution of the limit equation (6) and a scaled solution to the *linear* DK problem with *modified* parameters:

$$p = p_0(A, t) + C^* \frac{P_{DK}(\lambda^*, \delta_0^*, t - t_{ph})}{P_{DK}(\lambda^*, \delta_0^*, \infty)}, \tag{13}$$

This conjecture is well confirmed by numerical analysis; the numerical simulations show that one can always find $A$, $C^*$, $\lambda^*$, and $t_{ph}$ such that the function (13) accurately fits the numerical solution of the exact equation for the molecular state probability (4). Furthermore, the simulations indicate that there is no need to modify the detuning parameter $\delta_0$. It also turns out that we may put $t_{ph} = 0$. Further numerical analysis shows that the absolute error of formula (13) is less than $3 \times 10^{-6}$ for the final transition probability. For arbitrary times, its absolute error is commonly of the order of $10^{-5} - 10^{-4}$, and for the points of the first few maxima and minima of the function $p(t)$ (at certain values of the input parameters $\lambda$ and $\delta_0$) the deviation may increase up to $10^{-3}$.

Examining the role of the two terms in the approximate expression for the molecular state probability (13), we see that the first term, being a solution of the nonlinear equation (6), effectively describes the process of the molecule formation while the second one, being the scaled solution to the *linear* DK problem, describes the oscillations which arise some time after the system has passed through the resonance (see Fig. 4). From this, one can conclude that in the strong coupling limit the dynamics of the atom-molecule conversion effectively consists of the nonlinear resonance crossing followed by atom-molecular coherent oscillations that are principally of linear nature. The possibility to make such decomposition is not trivial since the governing set of equations (1) is essentially nonlinear.

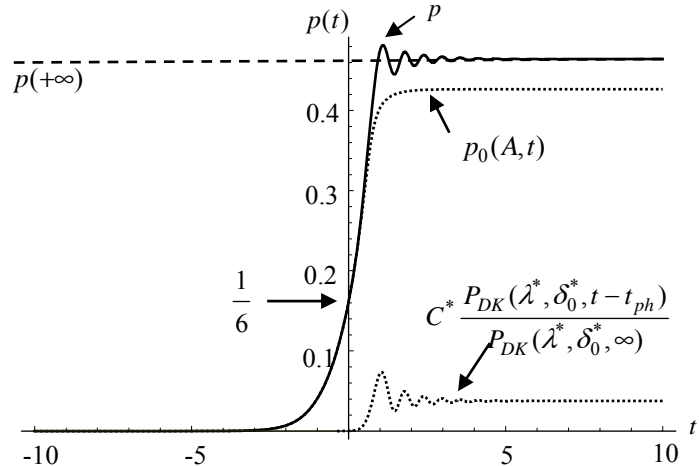

Fig. 4. Molecular state probability $p$, the limit solution $p_0$ determined from Eq. (6), and the scaled solution to the linear DK problem with modified parameters ($\lambda = 49$ and $\delta_0 = 5.5$).

One of the essential virtues of Eq. (13) is that it provides a simple expression for the final transition probability written in terms of the parameters $A$ and $C^*$:

$$p(+\infty) = p_0(+\infty) + C^*. \tag{14}$$

To study the asymptotic behavior of the fitting parameters $A$ and $C^*$ in the limit $\lambda \to \infty$, we substitute the trial function (13) into the exact equation for the molecular state probability (4) and consider the behavior of the remainder

$$R = \left(\frac{d}{dz} - 2\csc(2z)\right)\left\{4[\lambda(1-3p_0) - \lambda^*]C^* \frac{p_{DK}(\lambda^*, z(t))}{p_{DK}(\lambda^*, z(+\infty))} + C^* \frac{2\lambda^*}{p_{DK}(\lambda^*, z(+\infty))}\right. \tag{15}$$
$$\left. + (p_{0zz} - A) - 6\lambda C^{*2} \frac{p_{DK}^2(\lambda^*, z(t))}{p_{DK}^2(\lambda^*, z(+\infty))}\right\}.$$

It is seen that for $\lambda \gg 1$ the first term of the remainder is highly suppressed if we choose the fitting parameter $\lambda^*$ as

$$\lambda^* = \lambda(1 - 3p_0(+\infty)). \tag{16}$$

Then taking into account the value of $p_0(+\infty)$ [see Eqs. (8)-(9)], we have:

$$\lambda^* = -\frac{\lambda}{2} + 3\sqrt{\frac{A\lambda}{2}}. \tag{17}$$

Hence, we conclude that for $\lambda \gg 1$, $\lambda^*$ is a (large) *negative* parameter.

Regarding the two last terms of Eq. (15), one should minimize them with respect to the parameter $C^*$. This implies the condition

$$\frac{\partial(R/C^*)}{\partial C^*} = \left(\frac{d}{dz} - 2\csc(2z)\right)\left(-\frac{1}{C^{*2}}(p_0'' - A) - 6\lambda \frac{P_{DK}^2(\lambda^*, z(t))}{P_{DK}^2(\lambda^*, z(+\infty))}\right) = 0. \tag{18}$$

Since the last term is proportional to (large) $\lambda$ and $P_{DK}(\lambda^*, t)$ is an increasing function of time, the "worst" point is $t = +\infty$. Hence, we look for a minimization at $t = +\infty$. This immediately leads to the following value for $C^*$:

$$C^* = \sqrt{\frac{A}{6\lambda}}. \tag{19}$$

Thus, we have constructed approximate expressions for the fitting parameters $C^*$ and $\lambda^*$ in the case of large values of the peak laser field intensity and moderate values of the sweep rate through the resonance ($\lambda \gg 1$, $1 < \delta_0 < \sqrt{\lambda}$). Note that the parameters $C^*$ and $\lambda^*$ still depend on of the fitting parameter $A$.

## 3. Conclusion

The strong interaction limit of the nonlinear DK problem for coherent molecule formation in an atomic BEC via two-mode one-color photoassociation or a sweep across a Feshbach resonance has been examined in the mean-field approximation, for the case when the passage through the resonance is fast enough. In the case of photoassociation, the considered interaction regime models a situation when the peak laser field intensity is large, and the laser, being far detuned in the beginning of the interaction, passes, quasi-linearly in time, through the resonance, and eventually becomes again far detuned ($\lambda > 1$, $1 < \delta_0 < \sqrt{\lambda}$).

We have shown that the approximate expression for the molecular state probability can be effectively represented as a sum of two distinct terms (see Eq. (13)). The first term is defined as a solution of the limit first-order *nonlinear* differential equation (6) while the second one is a scaled solution to the *linear* DK problem with modified parameters. Examining the role of the different terms one can conclude that in the strong coupling limit the dynamics of the atom-molecule conversion consists of the process of nonlinear resonance crossing followed by atom-molecular coherent oscillations that are principally of linear nature. The possibility to make such decomposition is not trivial since the governing set of equations (1) is essentially nonlinear.

The constructed solution incorporates three auxiliary fitting parameters, $A$, $C^*$, and $\lambda^*$, the appropriate choice of which ensures that the constructed approximation describes the molecule formation process with high accuracy; absolute error is less than $3 \times 10^{-6}$ for the final transition probability. For arbitrary time points, its absolute error is commonly of the order of $10^{-5} \div 10^{-4}$, and for the points of the first few maxima and minima of the function $p(t)$ (at certain values of the input parameters $\lambda$ and $\delta_0$) it may increase up to $10^{-3}$.

I am indebted to professor Artur Ishkhanyan for his invaluable input during the whole research.

This research has been conducted in the scope of the International Associated Laboratory IRMAS. The work was supported by the Armenian National Science and Education Fund (ANSEF Grant No. 2009-PS-1692), the French Embassy in Armenia (Grant No. 2007-3849 Boursière du Gouvernement Français, Ecole Doctorale Lorraine de Chimie et Physique Moléculaires, Federal Agency for Education of Russian Federation and Armenian "Luys" Educational Foundation.

**References**

1. M.H. Anderson, J.R. Ensher, M.R. Matthews, C.E. Wieman, and E.A. Cornell, Science **269**, 198 (1995); K.B. Davis, M.-O. Mewes, M.R. Andrews, N.J. van Druten, D.S. Durfee, D.M. Kurn, and W. Ketterle, Phys. Rev. Lett. **75**, 3969 (1995); C.C. Bradley, C.A. Sackett, J.J. Tollett, and R.G. Hulet, *ibid.* **75**, 1687 (1995).
2. W.D. Phillips, Rev. Mod. Phys. **70**, 721 (1998); C.N. Cohen-Tannoudji, Rev. Mod. Phys. **70**, 707 (1998); S. Chu, Rev. Mod. Phys. **70**, 685 (1998).
3. H.R. Thorsheim, J.Weiner, and P.S. Julienne, Phys. Rev. Lett. **58**, 2420 (1987); R. Napolitano, J. Weiner, C.J. Williams, and P.S. Julienne, *ibid.* **73**, 1352 (1994).
4. W.C. Stwalley, Phys. Rev. Lett. **37**, 1628 (1976); E. Tiesinga, B.J. Verhaar, and H.T.C. Stoof, Phys. Rev. A **47**, 4114 (1993).
5. E.A. Donley, N.R. Claussen, S.T. Thompson, and C.E. Wieman, Nature **417**, 529 (2002); S. Durr, T. Volz, A. Marte, and G. Rempe, Phys. Rev. Lett. **92**, 020406 (2004); K. Xu, T. Mukaiyama, J.R. Abo-Shaeer, J.K. Chin, D.E. Miller, and W. Ketterle, *ibid.* **91**, 210402 (2003); J. Herbig, T. Kraemer, M. Mark, T. Weber, C. Chin, H.-C. Ngerl, and R. Grimm, Science **301**, 1510 (2003).
6. C.A. Regal, C. Ticknor, J.L. Bohn, and D.S. Jin, Nature **424**, 47 (2003); K.E. Strecker, G.B. Partridge, and R.G. Hulet, Phys. Rev. Lett. **91**, 080406 (2003); S. Jochim *et al.*, *ibid.* **91**, 240402 (2003); J. Cubizolles *et al.*, *ibid.* **91**, 240401 (2003). M. Greiner, C.A. Regal, and D.S. Jin, Nature **426**, 537 (2003);
7. M.W. Zwierlein, C.A. Stan, C.H. Schunck, S.M.F. Raupach, S. Gupta, Z. Hadzibabic, and W. Ketterle, Phys. Rev. Lett. **91**, 250401 (2003); S. Jochim, M. Bartenstein, A. Altmeyer, G. Hendl, S. Riedl, C. Chin, J. Hecker Denschlag, Science **302**, 2101 (2003).
8. A. Ishkhanyan, J. Javanainen, and H. Nakamura, J. Phys. A **38**, 3505 (2005).
9. A. Ishkhanyan, J. Javanainen, and H. Nakamura, J. Phys. A **39**, 14887 (2006).
10. R. Sokhoyan, H. Azizbekyan, C. Leroy, and A. Ishkhanyan, e-print arXiv:0909.0625 (2009).
11. A. Ishkhanyan, B. Joulakian, and K.-A. Suominen, J. Phys. B **42**, 221002 (2009).
12. R.A. Barankov and L.S. Levitov, arXiv:cond-mat/0506323v1 (2005).
13. E. Altman and A. Vishwanath, Phys. Rev. Lett. **95**, 110404 (2005).
14. I. Tikhonenkov, E. Pazy, Y.B. Band, M. Fleischhauer, and A. Vardi, Phys. Rev. A **73**, 043605 (2006).
15. B.E. Dobrescu and V.L. Pokrovsky, Phys. Lett. A **350**, 15 (2006).
16. A. Altland and V. Gurarie, Phys. Rev. Lett. **100**, 063602 (2008).
17. A. Altland, V. Gurarie, T. Kriecherbauer, and A. Polkovnikov, Phys. Rev. A **79**, 042703 (2009).
18. A.P. Itin, A.A. Vasiliev, G. Krishna, and S. Watanabe, Physica D **232**, 108 (2007).
19. A.P. Itin and P. Törmä, e-print arXiv:0901.4778 (2009).
20. R. Sokhoyan, D. Melikdzhanian, C. Leroy, H.-R. Jauslin, and A. Ishkhanyan, e-print arXiv:0910.3061 (submitted to Physica D) (2009).
21. L.D. Landau, Phys. Z. Sowjetunion **2**, 46 (1932); C. Zener, Proc. R. Soc. London, Ser. A **137**, 696 (1932).
22. N. Demkov and M. Kunike, Vestn. Leningr. Univ. Fis. Khim. **16**, 39 (1969); K.-A. Suominen and B.M. Garraway, Phys. Rev. A, **45**, 374 (1992).
23. R. Sokhoyan, H. Azizbekyan, C. Leroy, and A. Ishkhanyan, J. Contemp. Phys. (Armenian Ac. Sci.) **44**(6), 272 (2009).
24. R. Sokhoyan, B. Joulakian and A. Ishkhanyan, J. Contemp. Phys. (Armenian Ac. Sci.) **41** (3), 1 (2006).
25. A. Ishkhanyan, B. Joulakian, and K.-A. Suominen, Eur. Phys. J. D **48**, 397 (2008).
26. R. Sokhoyan, J. Contemp. Phys. (Armenian Ac. Sci.) **45** (2), 79 (2010).

27. J. Javanainen and M. Mackie, Phys. Rev. A **59**, R3186 (1999); M. Kostrun, M. Mackie, R. Cote and J. Javanainen, Phys. Rev. A **62**, 063616 (2000); M. Mackie and J. Javanainen, Phys. Rev. A **60**, 3174 (1999).
28. P.D. Drummond, K.V. Kheruntsyan, and H. He, Phys. Rev. Lett. **81**, 3055 (1998); D.J. Heinzen, R. Wynar, P.D. Drummond, and K.V. Kheruntsyan, Phys. Rev. Lett. **84**, 5029 (2000).